\documentclass[manuscript]{aastex}

\slugcomment{To be submitted to the Astronomical Journal.}

\shorttitle{Distance-Limited Imaging Survey of Sub-Stellar Companions}
\shortauthors{Carson et al.}

\begin{document}


\title{A Distance-Limited Imaging Survey of Sub-Stellar Companions to Solar Neighborhood Stars}


\author{Joseph C. Carson} 
\affil{Max-Planck-Institut f$\ddot{u}$r Astronomie, K$\ddot{o}$nigstuhl 17,
69117 Heidelberg, Germany}


\author{Kyle D. Hiner\altaffilmark{1},Gregorio G. Villar III, Michael G. Blaschak, Alexander L. Rudolph}
\affil{Physics Department,
California State Polytechnic University, Pomona,
3801 W. Temple Ave.,
Pomona, CA 91768, U.S.A. }

\and

\author{Karl R. Stapelfeldt}
\affil{Jet Propulsion Laboratory, 4800 Oak Grove Dr., Pasadena, CA 91109,
  U.S. A.}

\altaffiltext{1}{present address: Department of Physics and Astronomy,
University of California, Riverside,
900 University Ave.,
Riverside, CA 92507, U.S.A.}




\begin{abstract}
We report techniques and results of a Palomar 200-inch (5 m) adaptive optics
imaging survey of sub-stellar companions to solar-type stars.  The survey
consists of $K_{s}$ coronagraphic observations of 21 FGK dwarfs out to 20 pc
(median distance $\sim$17 pc).  At
1\arcsec separation (17 projected AU) from a typical target system, the survey achieves median
sensitivities 7 mag fainter than the parent star.  In terms of companion
mass, that corresponds to sensitivities of 50$M_{J}$ (1 Gyr), 70$M_{J}$
(solar age), and 75$M_{J}$ (10 Gyr), using the evolutionary models of Baraffe
and colleagues.  Using common proper motion to distinguish companions from
field stars, we find that no system shows positive evidence of a previously unknown substellar companion (searchable separation $\sim$20-250 projected AU at the median target distance).
\end{abstract}


\keywords{methods: data analysis --— stars: low-mass, brown dwarfs —--
  surveys --— techniques: high angular resolution  }



\section{Introduction}

    The discovery of the brown dwarf GJ 229B (Nakajima et al. 1995) heralded a
    stream of direct detections of substellar objects.  In particular, field
    surveys like the Sloan Digital Sky Survey (SDSS; Gunn \& Weinberg 1995),
    the Two Micron All Sky Survey (2MASS; Skrutskie et al. 1997), and the Deep
    Near Infrared Survey (DENIS; Epchtein 1997) helped raise the number of
    brown dwarf (L and T type) identifications today to over 600 \citep{gel08}.  However the number of
brown dwarfs identified as companions to main sequence stars remain few.  At the time of
this writing, there are only a handful of brown dwarfs confirmed as companions to main sequence stars.  Brown dwarfs that are part of stellar systems are particularly interesting because they often yield insights into brown dwarf and planet formation around stars.  For instance, statistics on the frequency of brown dwarf companions may shed light on the differences between planet, brown dwarf, and star formation.  And unlike the case of discovered field brown dwarfs, the presence of a central star often reveals additional information on the presumably co-evolved brown dwarf --– information like distance, metallicity, and age.    

     A number of high-contrast surveys have attempted to improve our knowledge of
     the moderate to wide separation (40AU to a few hundred AU) substellar
     companion population around stars.  For example, Biller et al. (2007) and
     Metchev \& Hillenbrand (2004) each surveyed samples (45 targets, 101
     targets) of young (age $\lesssim$250 Myr, $\lesssim$400 Myrs), relatively
     nearby ($\lesssim$50 pc, $\lesssim$160 pc) stars using adaptive optics
     (AO) systems on the VLT and Palomar/Keck telescopes, respectively;
     Lowrance et al. (2005) used HST NICMOS to survey 45 young (median age
     $\sim$150 Myr), nearby (average distance $\sim$30 pc) stars for
     substellar companions; Lafreni$\grave{e}$re et al. (2007) used Gemini AO
     to observe 85 young (median age $\sim$150 Myr), nearby (average distance
     $\sim$22 pc) stars;  Carson et al. (2005, 2006) used Palomar AO to survey
     80 nearby (median distance $\sim$17 pc) stars with unknown ages.  These references
     represent some of the larger direct-imaging high-contrast
     surveys, but there are a number of other surveys as well.     

The observations described in this document largely provide an extension to Carson et al. (2005), although the new target list is focused more strongly on solar-type stars.  (The Carson et al. [2005, 2006] survey looked mostly at K and M-dwarfs.)  Most of the competing surveys (like Biller et al. 2007, Metchev \& Hillenbrand 2004, Lowrance et al. 2005, Lafreni$\grave{e}$re et al. 2007, and others) have focused on observing the youngest nearby stars.  While this allows for a maximal substellar-object self-luminosity (Baraffe et al. 2003), it inherently requires that one examines stars at somewhat larger distances, in order to achieve a large enough original sample to glean the youngest stars.  Surveying only younger stars also leads to selection biases, as certain types of stars lend themselves better to age determinations than others (Mamajek \& Hillenbrand 2008).
For our survey, we avoid age requirements in an effort to achieve a more uniform census of the substellar companion population around the nearest solar-type stars.

     While explorations of the substellar space around nearby FGK stars are
     scientifically interesting in their own right, they also provide
     important reconnaissance observations for the next generation of
     planet-search imaging surveys, like expected programs with Subaru
     (HiCIAO; Tamura et al. 2006), Palomar (Project 1640; Hinkley et al. 2008),
     VLT (SPHERE; Beuzit et al. 2008), Gemini (GPI; Macintosh et al. 2006), and potential space missions like
    Terrestrial Planet Finder Coronagraph
    (TPF-C)\footnote{http://planetquest.jpl.nasa.gov/TPF-C} and Terrestrial
    Planet Finder Interferometer (TPF-I)\footnote{http://planetquest.jpl.nasa.gov/TPF-I}/Darwin\footnote{http://www.esa.int/science/darwin}.  Information on the presence of brown
     dwarfs is important for these future planet surveys because the existence
     of an orbiting
     brown dwarf may affect the likelihood of there being a planet in that
     system.  Even for southern hemisphere future planet searches, whose
     targets may not overlap with this document's survey, the statistical
     information (from our suvey and others) on brown dwarf frequencies may
     help guide southern surveys' overall target selection strategy.  In
     addition to information on orbiting brown dwarfs, future surveys will also
     benefit from our survey's reports on discovered field objects close to
     nearby stars.  Such information will ensure that future surveys do not
     spend unnecessary time following up these field objects for common proper
     motion tests.  This information will also guide target selection, by
     providing information on field objects whose interfering light might
     adversely affect high-contrast sensitivities.  

     In the sections below we present techniques and results of our recently completed survey.  Section 2 presents our target sample.  Section 3 describes our observing techniques.  In section 4 we present the data analysis techniques we used for this survey.  In section 5 we summarize our survey sensitivities.  Section 6 describes our results.  We present our conclusions in section 7.

\section{Targets}
     Our target selection process had its origins in work being carried out at
     Jet Propulsion Laboratory to select potential candidates for the proposed
     TPF-C mission \citep{lev06}.  The main constraints for that selection process included
     FGK V spectral type, no known stellar multiplicity, closeness ($<$20
     pc), and visible brightness (V $\lesssim$ 7).  For optimal observations
     from Palomar Observatory, we set a declination lower limit of -5 $\deg$.
     We also removed any targets from our list that were already observed
     during the Cornell High-Order Adaptive Optics Survey
     (CHAOS; Carson et al. 2005) and other adaptive optics imaging programs,
     as determined by a standard literature search.  We confirmed systems'
     lack of multiplicity by on-telescope preliminary imaging.  One target,
     GJ 564, had a previously published brown dwarf binary companion (Potter
     et al. 2002).
     This fact eluded us in our initial literature search and
     preliminary imaging. Hence, we ended up observing this target and
     recording its current astrometry.  

Our final target list
     consisted of 21 main sequence stars.  This included 2
     F stars, 14 G stars, and 5 K stars.  All stars possessed
     well-characterized proper motion and parallax values as defined by
     \textit{Hipparcos} (Perryman et al. 1997).  As nearby stars, they
     typically possessed high proper motion (median target proper motion $\sim$
     600 mas yr$^{-1}$), thus facilitating
     an efficient common proper motion follow-up strategy for candidate
     companions.  A complete list of the target set is given in Table 1.     

\section{Observations}
\subsection{Coronagraphic Search Observations}


To conduct our survey, we used the PALAO system (Troy et al.
2000) and the accompanying PHARO science camera (Hayward
et al. 2001) installed on the Palomar 200 inch Hale Telescope.
PALAO provided us with the high resolution (FWHM typically $\sim$ 0.14$\arcsec$ in
$K_{s}$ and strehl ratio around 50\%) necessary for resolving close companions.
The accompanying PHARO science camera (wavelength sensitivity
1--–2.5 $\micron$ and plate scale 40 mas pixel$^{-1}$) provided us
with a coronagraphic imaging capability, along with a maximal field of
view ($\sim$30$\arcsec$).
Our general observing strategy was to align the coronagraphic
mask on a target star and take a series of $K_{s}$ exposures, being careful to
not saturate many pixels in the detector. (Occasionally, we saturated
at the edges of the coronagraphic mask, where high noise
levels already prevented any meaningful companion search.)  We chose the
$K_{s}$ band because it provides high strehl ratios and a favorable relative
flux between substellar companion and parent star.
We planned our individual exposure times and number of exposures per set to allow
for no more than $\sim$ 5 minutes (including
overheads) between any target frame and the nearest sky image. This helped ensure that sky conditions did not significantly
change between the target exposures and accompanying sky exposures.  The sky
exposures consisted of our taking, before and after the target sequence, and
with the same setup as the target sequence, dithered images of a nearby empty sky
region.  We spent a comparable amount of time on sky as we did on
target.  We repeated the process of sky-target-sky as many times as was
appropriate, with the
goal of being able to detect (at 5-sigma) an $\sim$ 18-mag ($K_{s}$) object at
 5$\arcsec$ separation from the primary star.  Once we completed the
target/sky sets, we inserted a neutral density filter in the optical path and
conducted dithered non-coronagraphic exposures of the target star.  These images
allowed us to characterize and record instrument and site observing
conditions.  Table 1 gives the relevant observing information for the
individual targets.  




\subsection{Common Proper Motion Observations}
For candidate companions detected in the previous procedures, we checked for
physical companionship by using common proper motion observations.  The nearby
stars we observed tend to have high proper motions (on the order of a few
hundred mas yr$^{-1}$).  The vast majority of false candidate companions are
background stars that tend to have very small proper motions compared to the
parent star.  Therefore, after recording our initial measurement, we waited
for a timespan long enough for the parent star to move a detectable distance
 from the original position. In practice, this observable motion ended up
 being a minimum of $\sim$ 3 pixels.  After taking the second epoch
 observations, we checked
whether the candidate
maintained the same position with respect to the parent star.  Target stars
re-observed to check for common proper motion included GJ 895.4, 159, 230,
1095, 1085,
56.5, and 788.  GJ 778 and 758 contained candidate companions, but were not
re-observed due to scheduling constraints (see discussion in Section 6).

\section{Data Analysis}
\subsection{Reducing Images}
We began our data reduction by median-combining each of the dithered sky
sets.  We then took each coronagraphed star image and subtracted the
median-combined sky taken closest in time to the star image.  (The typical
separation in time between target and sky image was $\sim$5.5 minutes.)  We
divided each of the sky-subtracted star images by a flat-field frame that we
created, using standard procedures, from dark-frames and the dithered sky sets.  We chose to use sky-flats, instead of the more conventional twilight or dome flats, in order to have a flat-field frame generated with an optical path most similar to that used with the target observation.  We felt that this was important because we have seen transmission features (which we suspect reside on PHARO's moveable optics)  that change their observable position (by a few pixels perhaps) over the course of the night.
 
 After performing the flat-fielding, we next median-combined each
sequence of coronagraphic star frames.  For this median combination, we used
the images' residual parent star flux (which leaked from around the
coronagraph) to realign any frames that may have shifted because of instrument
flexure.  (Typical re-alignments were $\lesssim$ 1 pixel.)  We concluded that in-software re-alignment produced better overall sensitivities than just throwing-out mis-aligned frames.   Next we applied a bad-pixel algorithm to remove suspicious pixels
(defined as any pixel deviating from the surrouding 8 pixels by $\geq$5
$\sigma$) and replace them with the median of their 8 neighbors.  

We next applied a Fourier filter to the resulting images to help remove
non-point-like features such as unwanted internal instrument reflection and
residual parent star flux.  The Fourier filter application entailed our
multiplying a Fourier-transformed version of the final image
 by an exponential transmission function that minimized lower frequency signals.   This Fourier filter, described in \citep{car05}, has been shown to improve S/N by $\sim$ 25\% for a typical PHARO high-contrast image.    
  Along with this S/N
improvement, the typical point spread function (PSF) FWHM decreased by about 10\% as a result of the
Fourier filter application.  

Finally, we investigated possibilities for taking advantage of the approximate symmetry of the coronagraphed PSF to self-subtract an inverted and/or rotated version of the PSF from the non-inverted image.  We ended up deciding against using this technique as the final improvement was either marginal or non-existent.

\subsection{Identifying Brown Dwarf Companions}

Our first step in identifying candidate brown dwarf companions was to
         individually inspect each final Fourier-filtered and non-Fourier-filtered
image for any potential companions.  By choosing to examine both Fourier-filtered 
         and non-Fourier-filtered final images, we effectively recognize that the
        filtering technique improves S/N in some
         instances and worsens it in others.  For instance, Fourier-filtering works best in regions with shallowly sloping unwanted signal, like regions with internal instrument reflection.  However, in other regions, the candidate companion S/N may suffer since the filtering always removes some true candidate companion flux.  

The Palomar AO characteristic PSF ``waffle pattern'' (see Figure 1) is often seen as undesirable by observers, as the pattern degrades the potential PSF sharpness.  While this is true in principle, the characteristic ``waffle pattern'', in practice, provided an important way for us to distinguish true celestial objects (which had a well-patterned four-cornered PSF) from statistical outliers in the image noise, which typically took on more arbitrary shapes.  Thus, it provided an important first step in candidate companion identification.

While individual inspection of waffle patterns proved useful, we chose to also
         use an automated detection system to deliver more quantifiable sensitivity levels.
Our automated algorithm operated by centering on every other
  pixel in the final reduced Fourier-filtered and non-Fourier-filtered image
  and creating there a 0.$\arcsec$16 diameter flux aperture and
  1.$\arcsec$2-1.$\arcsec$6 diameter sky annulus.  After subtracting the background, the algorithm approximated a
  S/N level by dividing the measured aperture flux by the
  combined aperture flux Poisson noise and background noise;  it
  approximated background noise from the standard deviation of the sky annulus
  pixels. In
          the end it outputted a final array with a S/N
          value for each sampled pixel.  For each S/N map,
  there was also generated a map of measured background noise at each
  position (as estimated from the sky annuli).  This
  outputted noise map essentially reflected the ability of the
  algorithm to detect (at a given thresh-hold S/N level)
  different brightness objects according to position on the array.

After generating maps for a given image, the program selected
          the S/N map pixel with the highest value, using a
          minimum value of five.  It recorded the
          pixel position and then moved on to record the next highest
          S/N value greater than five.  After each
          detection, it voided a 0.$\arcsec$4 radius around the
          detected candidate object.  This procedure continued
          until there were no more positions with S/N
          values greater or equal to five.  (Of course, for many
          images, no positions possessed S/N levels
          greater than five.)  After the algorithm identified the candidate
          sources,
          we re-examined the final images to ensure that the
          algorithm had indeed detected a true source as opposed to a
          systematic effect.  Again, we searched for the Palomar adaptive
          optics signature ``waffle pattern'' to ensure a true physical
          source.  We also made comparisons to images taken at other
          sources to ensure that the feature was indeed unique
          to the target image.  In practice, we found that individual image inspection, by eye, produced the most thorough identification of candidate companions.  However, we felt that the automated detection was important as well, in order to provide quantifiable detection sensitivities, as well as a second-check for our visual inspections.

We acknowledge that
          the use of our automated detection routine has some
          drawbacks.  Notably, there are instances where the
          algorithm over-estimates noise levels.  For instance, close to the parent star PSF, the algorithm can
          mistake what may be a well-ordered parent star PSF
          slope for a random fluctuation in background noise.  
            Additionally, the algorithm may also over-estimate the noise close to field stars;  if a
          field star happens to fall in the sky annulus, the algorithm will
          determine that region to have
          excessively high background noise.  Thus, only the brightest
          candidate
          objects would be detected near these field star positions.  While these instances are not ideal, we conclude that is an acceptable compliment to our careful visual inspections. 
  In Section 5 we discuss how we may generate
          limiting magnitudes and brown dwarf mass limits from these
          algorithm-generated noise maps.

In cases where we positively  identified a potential brown dwarf companion, we next estimated its apparent $K_{s}$ magnitude by comparing its flux to the
        non-coronagraphic parent-star calibration images 
       and published 2MASS $K$-magnitudes (Skrutskie et al. 1997).  Resulting
       magnitudes are
       displayed in Table 2.  Once we established an apparent $K_{s}$
       magnitude, we derived a corresponding absolute $K_{s}$
       magnitude, assuming the candidate had a distance equal to the
       parent system.  Thanks to observational surveys such as
       \textit{Hipparcos} (Perryman et al. 1997), all of our parent stars had
       well-defined parallaxes and therefore distances.  With an
       approximate absolute $K_{s}$ magnitude in hand, we combined
       published brown dwarf observational data (Leggett et al. 2000,
       Leggett et al. 2002, Burgasser et al. 1999, Burgasser et
       al. 2000,  Burgasser et al. 2002, Burgasser et al. 2003, Geballe et al. 2002, Zapatero et al. 2002,
       Cuby et al. 1999, Tsvetanov et al. 2000, Strauss et al. 1999,
       and Nakajima et al. 1995)  with theoretical data from Baraffe et al. (2003) to extrapolate constraints on the object's mass.  An
       object whose potential mass fell within acceptable brown dwarf restrictions was
designated for common proper motion follow-up observations. 

    For our follow-up observations, we used Hipparcos published
     common proper motion values (Hipparcos catalogue; Perryman et
     al. 1997) to determine the expected movement of the parent
     system.  Since background and field stars are unlikely to possess
     proper motions identical to the parent system's, we used common
     proper motion as a strong support for a physical companionship.
     For candidate companions, we used the PSF central peak to identify position.  For the obscured parent star, we used the waffle-pattern four corners (which resided outside the opaque coronagraph spot) to create a well-defined cross-hairs that revealed the central position.  We could typically constrain the relative offset between parent star and candidate companion by fractions of a pixel, depending on S/N levels.
     Measuring the candidate companion's relative position over the
     two epochs, we were able to distinguish physical companionships
     from chance alignments.  We record positions in
     Table 2.

\section{ Survey Sensitivities}

\subsection{Determining Limiting Magnitudes}

     To quantify detection sensitivities from the
     algorithm-generated noise maps described
     in Section 4.2, we
     looked to determine the faintest detectable magnitude as a
     function of angular separation from each parent star.  We began
     by sampling each noise map (including those deriving from Fourier-filtered and non-Fourier-filtered images)
     and selecting, for each pixel, the smaller of the two
     noise values.  The resulting composite noise map array therefore
     reflected
 the best
     sensitivities from each of the two final images.     Figure 2 displays a
sample image sequence, where a Fourier-filtered
     and non-Fourier-filtered image are combined to create a composite noise
     map.

    For the  composite noise map, we next determined the median values
     in a series of concentric 0$\arcsec$.20-thick rings centered on
     the noise map center.  The median values therefore
     represented typical noise as a function of distance from
     the central star.  For each noise value, we then determined the
     minimum apparent $K_{s}$-magnitude where signal exceeded the
     combined Poisson noise and ring noise by a factor greater or
     equal to 5;  we were able to convert noise values (in units of detector
     counts) to $K_{s}$-magnitudes using parent star calibration data
     described in Section 3.1.  In Figure 3 we plot resulting measurements for
     median survey sensitivities (middle
     curve), the best 10\% of
     observations (lower curve), and the worst 10\% of
     observations (top curve).  Refer to Table 3 for a summary of minimum
     detectable magnitudes for each of the individual targets.

     Another commonly used statistic for describing sensitivities for
     high-contrast companion surveys is the limiting differential
     magnitude as a function of angular separation from the parent
     star.  In other words, how many times dimmer may a companion
     object be before we lose it in the parent star noise?  Figure
     4
     plots differential magnitudes for median survey sensitivities  as
     well as the best
     and worst 10\% of observations.

\subsection{Mass Sensitivities}
  Determining sensitivities according to companion mass is complicated
  by the fact that brown dwarfs of a given mass dim over time.
  Nonetheless, to get a general idea of detectable masses, we may
  assume different test ages and then use models by Baraffe et al. (2003) to
  transform our minimum detectable brightnesses into brown dwarf
  masses.   Figure
  5 shows a comparison of median sensitivities assuming 1 Gyr, solar
  age, and 10 Gyr target ages.

\section{Results}
     After conducting all of our data analysis, we concluded that zero
     systems showed positive evidence of a brown dwarf
     companion (that was not previously known).  We did re-detect the brown dwarf binary orbiting GJ 564,     discovered by Potter et al. (2002).     
  In total, we detected 48 field objects (including the binary brown dwarf)
  around 10 targets stars.  The GJ 778 and GJ 758 fields both contained
  candidate companions, but were not re-observed for common proper motion
  follow-up tests due to scheduling constraints.  2MASS data (Skrutskie et
  al. 1997) reports K-band field star densities of $\sim$12 stars and $\sim$6 stars per
  PHARO field of view, for the respective GJ 778 and GJ 758 star neighborhoods.  Given the
  relatively high field star densities for these regions, the chances of these
  candidates being field stars are high. 
In
  the end, however, we cannot confirm or reject that one of these candidates may be a
  brown dwarf companion. 
    Instead, we simply report the photometry
  and astrometry for the detected objects.  Table 2 presents the measurements
  for these objects   
   as well as all the other field objects detected in our survey.

\section{Discussion}
As mentioned in the previous paragraph, our survey found no evidence of
new brown dwarf companions, for orbital separations akin to
our own outer solar system.  However, even for targets with no candidate
companions,   we cannot rule out the
possibility that one or more new brown dwarfs exist around the targeted stars, even at the semi-major axes for
which our
survey is most sensitive.  For instance, a substellar companion near conjuction, in an orbit close to edge-on, may be impossible to resolve from the parent star PSF, regardless of the companion's luminosity or semi-major axis.  
  Furthermore, even for a bright brown dwarf with a face-on orbital
inclination, a brown dwarf's orbital eccentricity might lead to a range of
possible projected separations, which could lead to a null detection.

     Extracting rigorous companion statistics is therefore complicated by factors such as unknown orbital characteristics.  For example, if typical brown dwarf orbits are highly eccentric, the typical semi-major axis regime that our survey covers is most likely narrower than the \textit{projected} orbital separation we probe, as shown in orbital simulations presented in Carson et al. (2006).  Furthermore, extracting the companion fraction for a given substellar mass range is complicated by the fact that one must assume a system age in order to translate the survey sensitivity floor (in terms of $K_{s}$ mag) into a minimum detectable mass (see discussion in Section 5.2).  Since most of our stars have unknown ages, to extract a companion fraction, one must resort to a statistical inference of target star ages, using a method such as galactic birth models (like that used in Burgasser 2004) or stellar metallicity relations (like that used in Carson et al. 2006).  (Alternative age determination methods, such as those using Ca II emission, lithium abundance, and X-ray activity, provide poor constraints for target sets older than a couple hundred Myrs, and are therefore not useful for our applications.)  

The extractable companion frequency also depends on the relative mass function of substellar companions.  For example, even if we limit ourselves to a constrained mass range (like 20 $M_{J}$ to 40 $M_{J}$), the companion fraction uncertainties may depend sensitively on whether the majority of brown dwarfs resides near the lower boundary or the higher boundary of our mass range; if the majority resides near the lower mass boundary, there is a greater chance that our null result is due to limiting sensitivities, as opposed to a true lack of companions.  

It is possible for one to make educated assumptions for all of the aforementioned factors, and then run detailed Monte Carlo population simulations to conclude a companion fraction (e.g. Nielsen et al. 2008, Carson et al. 2006).  Published Monte Carlo population analyses have shown that a large target data set ($\gtrsim$ 60 stars) is typically required to provide meaningful results.  For instance, Carson et al. (2006), using an 80 star sample, concludes a brown dwarf companion (25-100 AU semi-major axis) fraction of 0\% to 9\%.  Nielsen et al. (2008) concludes, from a 60 star sample, a planet/brown-dwarf ($>$ 4 $M_{J}$) companion (20-100 AU semi-major axis) fraction of 0\% to 20\%.  Considering these relatively large uncertainties, and that our 21 star sample is significantly smaller than these other surveys, we believe that we cannot conclude meaningful companion fractions from our data set alone.  We postpone, for a future paper, a detailed Monte Carlo population analysis that combines several surveys' data sets to derive the most meaningful companion fractions.       

In addition to explorations of brown dwarf companions,
our survey also reported astrometry and photometry for all detected field objects.  The reporting of such objects may have
 potential benefits to future surveys,  by providing possible reference
 star candidates, preventing future brown dwarf candidate false identifications, and
 yielding data on objects whose interfering light might impede future
 planet-search observations.

\acknowledgments

We thank the Palomar Observatory staff for their support of these
observations.  We thank our anonymous referee for useful comments.  Part of the research conducted by J.C.C was supported by an
appointment to the NASA Postdoctoral Program at the Jet Propulsion Laboratory,
administered by Oak Ridge Associated Universities through a contract with NASA.



{\it Facilities:} \facility{Hale}

\clearpage



\begin{figure}
\plotone{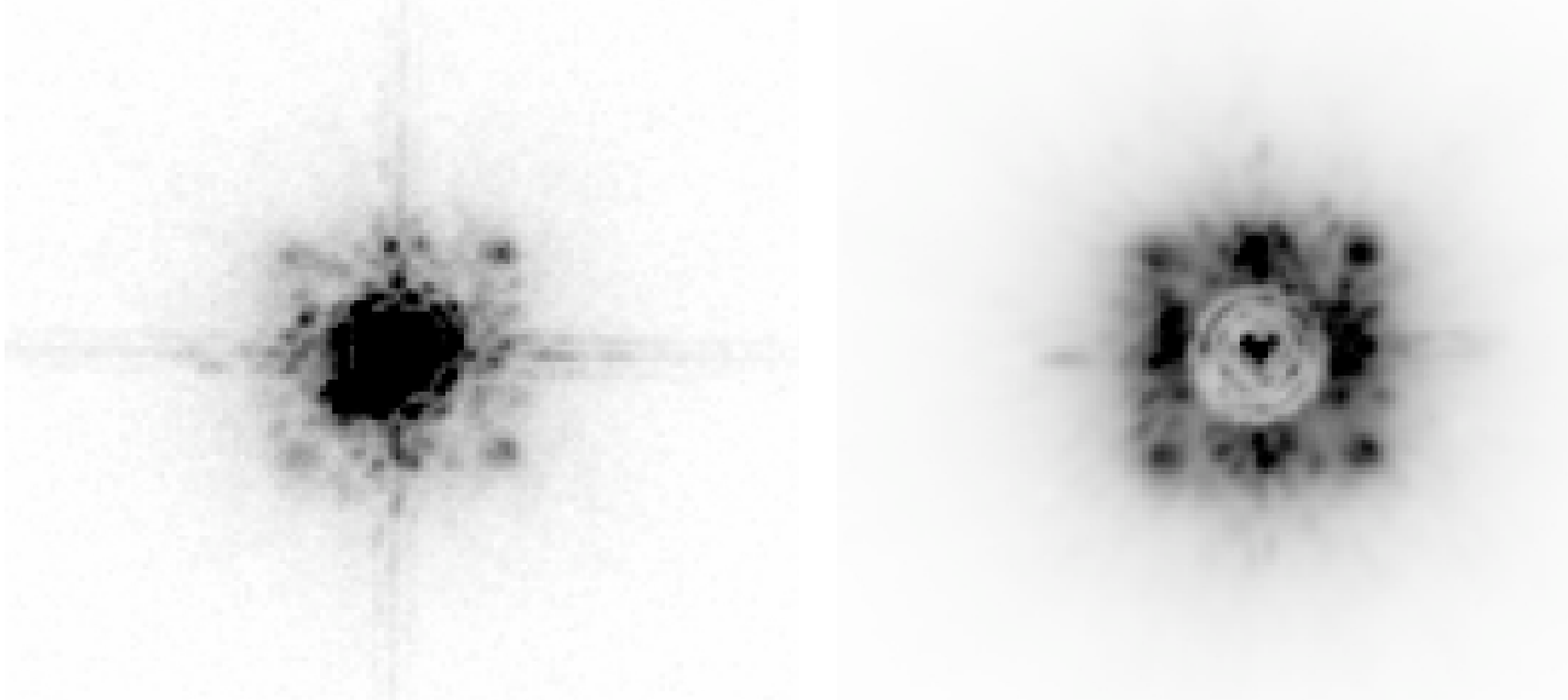}  
\caption{ Two example reduced $K_{s}$ images of GJ 230, taken in November,
            2007 using the Palomar 200-inch Adaptive Optics and accompanying
           PHARO science camera.  The image on the left is a non-coronagraphic
       GJ 230 image taken with a neutral density filter.  The slight
       elongation of the central PSF feature toward the lower left direction is an
       artifact of the neutral density filter, and does not occur in
       non-neutral-density-filter images.   The image on
   the right was taken with no neutral density filter, but with a  $0.\!''91$
   opaque spot positioned over the star.  The images illustrate             
  PALAO's characteristic AO-reconstructed PSF as seen in both                  
  coronagraphic and non-coronagraphic imaging.}
\end{figure}

\begin{figure}
\plotone{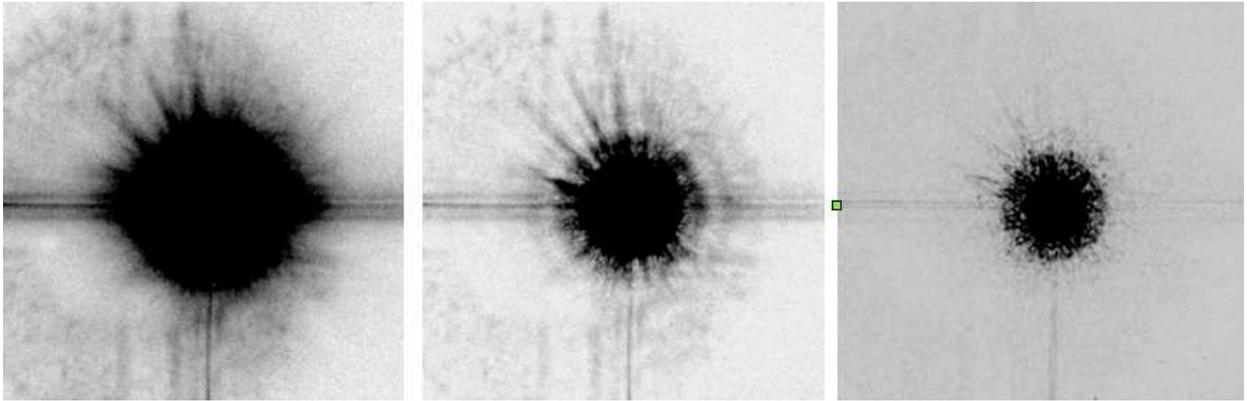} 
\caption{GJ 230 Final Images and Noise Map.  a) Left: GJ 230                  
  coronagraphic final image.  b) Center: GJ 230 coronagraphic,
  Fourier-filtered,  final   image.  c) Right: a composite noise map created from (a) and (b).  
The
  greyish, slightly offset ring segments around the star in (a) and (b) result from
  internal instrument reflection.
  }
\end{figure}

\begin{figure}
\plotone{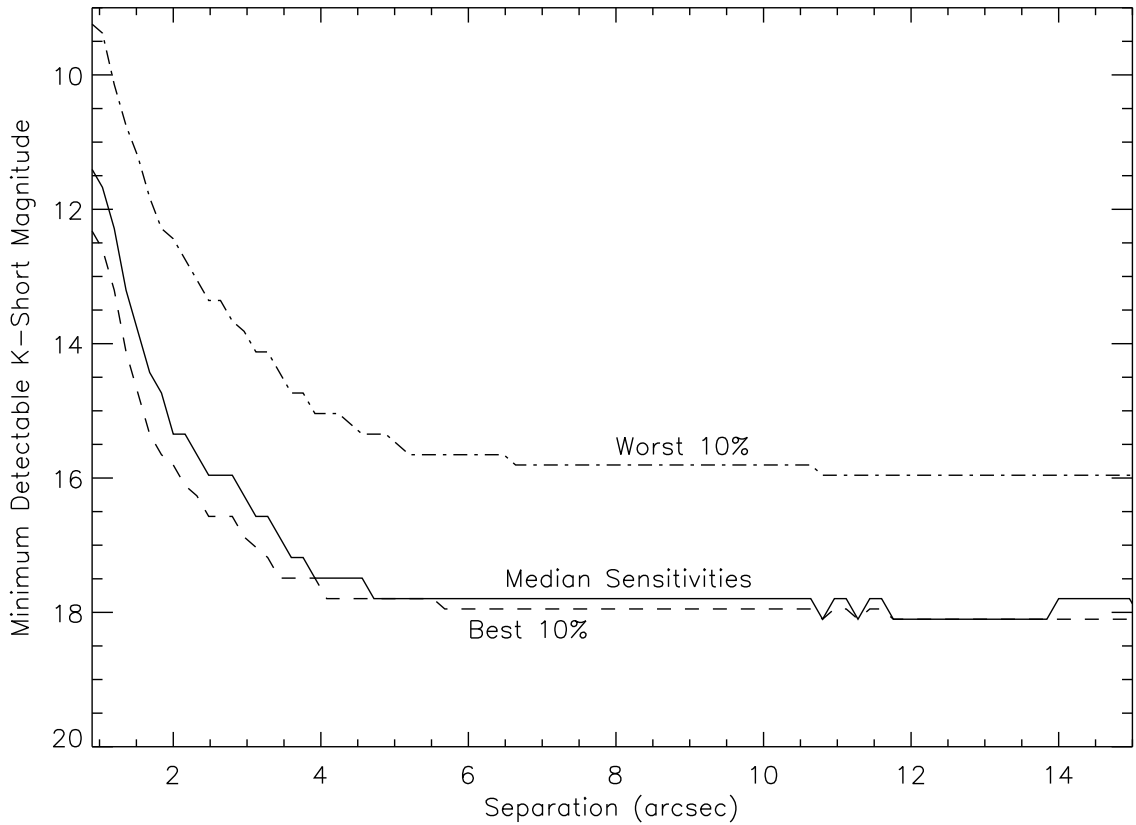}  
\caption{$K_{s}$-band sensitivity curves displaying limiting magnitude         
  as a function of separation from the parent star.  The top curve             
  represents the median sensitivities for the  worst 10\% of             
  observations. The middle curve represents median survey                      
  sensitivities.  The                                                          
  bottom curve represents median sensitivities for the best 10\% of            observations.   ``Best 10\%'' and ``Worst 10\%'' are defined by a
  combination
 of parent star                                       brightness, seeing conditions, and adaptive optics performance.  All minimum
 magnitudes correspond to 5-sigma detections.}
\end{figure}

\begin{figure}
\plotone{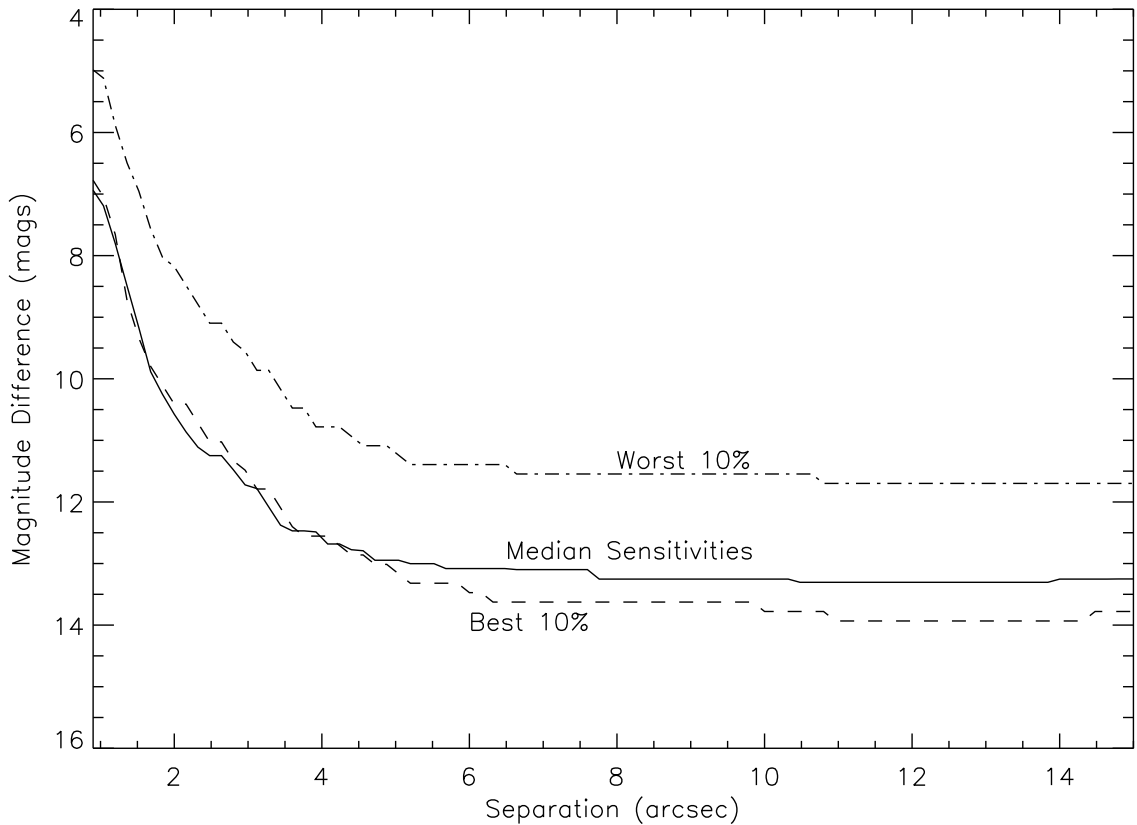} 
\caption{$K_{s}$-band sensitivity curves displaying limiting
  differential magnitude ($K_{s}$-companion minus $K_{s}$-parent) as a
  function of separation from the parent star.  The  middle curve
  represents median survey sensitivities.  The top curve represents
  median sensitivities for the worst 10\% of our data.  The
  bottom curve shows the best 10\%.  Limits represent 5-sigma detections.}
\end{figure}

\begin{figure}
\plotone{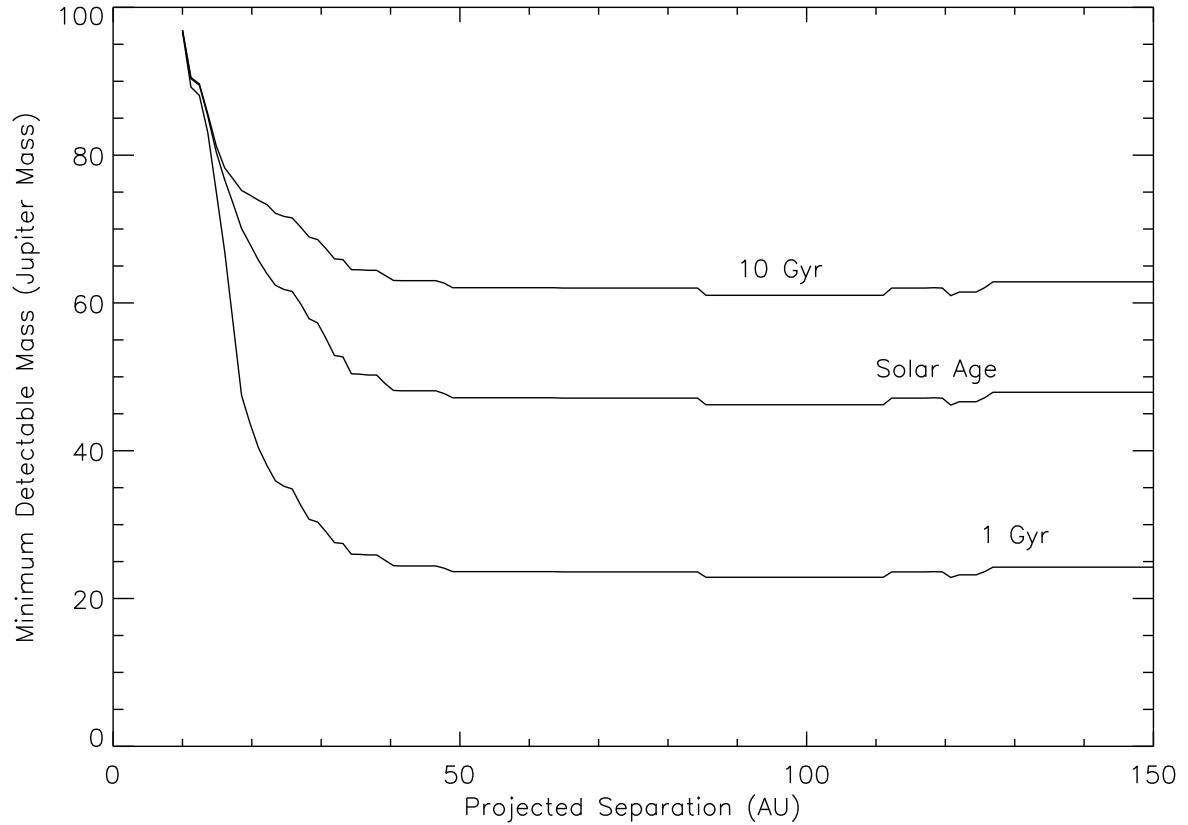}  
\caption{  Minimum detectable mass as a function of projected separation
   for median target sensitivities and distances.  We compare
  results for assumed 1 Gyr, solar age, and 10 Gyr targets, using
  evolutionary models by Baraffe et al. (2003).  We derived these
  curves by first plotting minimum detectable K-short magnitude
  versus arcsecond separation for all targets.  Next we
  median-combined the K-short curves to derive typical sensitivities (see Figure 3, middle curve).  We transformed these K-short magnitudes
  into masses using Hipparcos distances and Baraffe et al. (2003)
  evolutionary models.  Finally, we transformed our
  arcsecond axis to a projected AU separation using the median target distance.}     
\end{figure}
 
\begin{deluxetable}{lllllcc}
\tabletypesize{\scriptsize}
\rotate
\tablecaption{Nearby Star Target List}
\tablewidth{0pt}
\tablehead{
\colhead{Parallax} &    \colhead{\hspace{2cm} Proper Motion} & &
\colhead{V} & & Dates of Coronagraphic & Net Exposure \\
\colhead{(mas)} &
\colhead{RA (mas yr$^{-1}$)} & \colhead{Dec (mas yr$^{-1}$)} &
\colhead{(mag)} & \colhead{Name} & Observations & Time (sec) \\
}
\startdata
59.31 &  \hspace{1.05cm}   \hspace{-0.2cm} 1106.1 &       113.0 &
6.76 &  GJ 895.4 & 2005 Dec; 2006 May; 2007 Nov & 4916 \\

52.00&    \hspace{1.05cm}   \hspace{-0.2cm} 151.2& \hspace{-0.2cm}
-252.0&       5.38 &   GJ 159 & 2006 Dec; 2007 Nov & 581 \\

55.20&   \hspace{1.05cm}   \hspace{-0.2cm} 78.1&        -297.1&
6.43&   GJ 230 & 2005 Dec; 2007 Nov & 2149 \\

59.31&    \hspace{1.05cm}  29.4&        -186.1& 5.54&   GJ 1095
& 2006 Dec; 2007 Nov & 586 \\

59.46&     \hspace{1.05cm}  -34.1& \hspace{-0.2cm} -34.45&
7.17&   GJ 56.5 & 2006 Dec; 2007 Nov & 581 \\

64.25&     \hspace{1.05cm}  \hspace{-0.2cm} 62.4&       -230.7&
5.97&   GJ 1085 & 2006 Dec; 2007 Nov & 595 \\

59.52&     \hspace{1.05cm}  \hspace{-0.2cm} -171.2&
\hspace{-0.2cm}   -1164.2&   6.97&    GJ 295 & 2006 Dec & 340 \\

57.57&
\hspace{1.05cm} \hspace{-0.2cm} -359.8&       \hspace{-0.2cm}
139.3&       5.95&   GJ 484 & 2006 Dec & 283 \\

56.92&      \hspace{1.05cm} 468.5& 296.8& 5.91&   GJ 788 &
2006 May; 2006 Dec & 3054 \\

64.71&   \hspace{1.05cm}    -122.3&        \hspace{-0.2cm} -103.3&
6.76&   GJ 227 & 2005 Dec; 2006 Dec & 1982 \\

52.25&     \hspace{1.05cm}  \hspace{-0.2cm} -191.1& -115.4& 5.95&
GJ 334.2 & 2006 Dec & 297\\

71.04&    \hspace{1.05cm}  -315.9& \hspace{-0.2cm} 55.2&
5.03&   GJ 407 & 2005 Dec; 2006 May & 2046\\

56.82&    \hspace{1.05cm}   223.7& \hspace{-0.2cm} -477.5&
6.25&   GJ 547 & 2006 May & 1113\\

55.11&     \hspace{1.05cm}  132.5&  -298.4&  6.61&   GJ 614 & 2006
May & 1338\\

64.54&     \hspace{1.05cm}  \hspace{-0.2cm} 82.0& 162.9& 6.37&
GJ 758 & 2006 May & 361\\

67.14&    \hspace{1.05cm}   \hspace{-0.2cm} -529.2&        -428.9&
5.37&   GJ 376 & 2006 May & 2028\\

55.73&     \hspace{1.05cm}  144.7&  \hspace{-0.2cm} 32.4&
5.86&   GJ 564 & 2006 May & 487\\

69.61&    \hspace{1.05cm}   \hspace{-0.2cm} -571.2&
\hspace{-0.2cm} 52.6&        6.66&   GJ 611 & 2006 May & 476\\

55.37&    \hspace{1.05cm}  123.5&        \hspace{-0.2cm} 854.7&
6.76&   GJ 651 & 2006 May & 1427\\

64.17&    \hspace{1.05cm}  \hspace{-0.2cm} -1002.8&  \hspace{-0.2cm}
-912.6&  7.28&   GJ 778 & 2006 May & 1517\\

70.07&     \hspace{1.05cm}  \hspace{-0.2cm} -27.7&        87.9&
5.63&   GJ 311 & 2006 Dec & 297\\
\enddata
\tablecomments{Parallax, proper motion, and V-magnitude are from Hipparcos
  (Perryman et al. 1997).  All names follow the Gliese
catalog system (Gliese \& Jahreiss 1991).}

\end{deluxetable}

\begin{deluxetable}{lllll}
\tabletypesize{\scriptsize}
\tablecaption{Detected Field Objects}
\tablewidth{0pt}
\tablehead{
 & & & \colhead{Observation} & \\
\colhead{Parent Field}  &\colhead{$\rho$ (arcsec)\tablenotemark{a}}
&\colhead{$\theta$ (deg)\tablenotemark{b}} &\colhead{Date}
&\colhead{$K_{s}$ (mag)}
}
\startdata
GJ 895.4 & 9.619 $\pm$ 0.006 & 120.7 $\pm$ 0.3 & 2007 Nov & 15.53 $\pm$
0.16 \\
GJ 895.4 & 5.375 $\pm$ 0.008 & 287.1 $\pm$ 0.3 & 2007 Nov & 16.94 $\pm$
0.17 \\
GJ 895.4 & 12.992 $\pm$ 0.005 & 308.7 $\pm$ 0.3 & 2007 Nov & 17.11 $\pm$
0.12 \\
GJ 895.4 & 12.724 $\pm$ 0.004 & 328.9 $\pm$ 0.3 & 2007 Nov & 17.67 $\pm$
0.11 \\
GJ 895.4 & 11.629 $\pm$ 0.008 & 342.30 $\pm$ 0.3 & 2007 Nov & 17.13 $\pm$
0.07 \\
GJ 895.4 & 11.544 $\pm$ 0.007 & 341.7 $\pm$ 0.3 & 2007 Nov & 17.35 $\pm$
0.09 \\
GJ 895.4 & 9.771 $\pm$ 0.006 & 347.9 $\pm$ 0.3 & 2007 Nov & 18.24 $\pm$
0.12 \\
GJ 895.4 & 6.572 $\pm$ 0.100 & 338.4 $\pm$ 0.3 & 2007 Nov & 19.13 $\pm$
0.13 \\
GJ 895.4 & 16.279 $\pm$ 0.009 & 315.1 $\pm$ 0.3 & 2007 Nov & 18.24 $\pm$
0.11 \\
GJ 895.4 & 3.370 $\pm$ 0.004 & 118.0 $\pm$ 0.4 & 2007 Nov & 16.81 $\pm$
0.08 \\
GJ 159 & 15.747 $\pm$ 0.013 & 310.8 $\pm$ 0.3 & 2006 Dec & 16.89 $\pm$
0.02 \\
GJ 230 & 16.508 $\pm$ 0.007 & 321.8 $\pm$ 0.3 & 2007 Nov & 18.58 $\pm$
0.04 \\
GJ 1085 & 10.600 $\pm$ 0.007 & 148.1 $\pm$ 0.3 & 2006 Dec & 17.12 $\pm$
0.08 \\
GJ 1085 & 13.892 $\pm$ 0.025 & 226.2 $\pm$ 0.3 & 2006 Dec & 17.63 $\pm$
0.12 \\
GJ 1095 & 10.459 $\pm$ 0.006 & 84.4 $\pm$ 0.3 & 2006 Dec & 15.79 $\pm$
0.03 \\
GJ 1095 & 9.491 $\pm$ 0.003 & 87.5 $\pm$ 0.3 & 2006 Dec & 18.91 $\pm$
0.34 \\
GJ 56.5 & 9.510 $\pm$ 0.023 & 130.0 $\pm$ 0.3 & 2006 Dec & 16.97 $\pm$
0.06 \\
GJ 788 & 7.263 $\pm$ 0.008 & 119.0 $\pm$ 0.3 & 2006 May & 17.78 $\pm$
0.20 \\
GJ 788 & 7.096 $\pm$ 0.009 & 119.0 $\pm$ 0.3 & 2006 May & 18.09 $\pm$
0.2 \\
GJ 564\tablenotemark{*} & 2.600 $\pm$ 0.019 & 101.3 $\pm$ 0.4 & 2006 May & -- \\
GJ 564\tablenotemark{*} & 2.600 $\pm$ 0.012 & 102.6 $\pm$ 0.4 & 2006 May & -- \\
GJ 758 & 11.078 $\pm$ 0.009 & 57.0 $\pm$ 0.3 & 2006 May & 13.12 $\pm$ 0.03 \\
GJ 758 & 7.216 $\pm$ 0.006 & 358.8 $\pm$ 0.3 & 2006 May & 13.38 $\pm$ 0.04 \\
GJ 758 & 10.396 $\pm$ 0.003 & 193.4 $\pm$ 0.3 & 2006 May & 8.43 $\pm$ 0.04 \\
GJ 758 & 10.279 $\pm$ 0.003 & 210.9 $\pm$ 0.3 & 2006 May & 12.88 $\pm$ 0.03 \\
GJ 758 & 14.297 $\pm$ 0.015 & 135.2 $\pm$ 0.3 & 2006 May & 14.12 $\pm$ 0.03 \\
GJ 758 & 11.131 $\pm$ 0.012 & 319.2 $\pm$ 0.3 & 2006 May & 15.09 $\pm$ 0.10 \\
GJ 758 & 15.862 $\pm$ 0.010 & 235.0 $\pm$ 0.3 & 2006 May & 13.37 $\pm$ 0.04 \\
GJ 778 & 9.778 $\pm$ 0.007 & 101.8 $\pm$ 0.3 & 2006 May & 17.61 $\pm$ 0.02 \\
GJ 778 & 8.143 $\pm$ 0.005 & 87.1 $\pm$ 0.3 & 2006 May & 16.00 $\pm$ 0.02 \\
GJ 778 & 6.421 $\pm$ 0.007 & 81.8 $\pm$ 0.3 & 2006 May & 16.63 $\pm$ 0.04 \\
GJ 778 & 8.222 $\pm$ 0.008 & 310.3 $\pm$ 0.3 & 2006 May & 14.81 $\pm$ 0.02 \\
GJ 778 & 8.857 $\pm$ 0.010 & 301.3 $\pm$ 0.3 & 2006 May & 13.52 $\pm$ 0.03 \\
GJ 778 & 11.111 $\pm$ 0.006 & 263.7 $\pm$ 0.3 & 2006 May & 13.61 $\pm$ 0.03 \\
GJ 778 & 6.151 $\pm$ 0.008 & 227.3 $\pm$ 0.3 & 2006 May & 15.21 $\pm$ 0.03 \\
GJ 778 & 14.726 $\pm$ 0.008 & 303.2 $\pm$ 0.3 & 2006 May & 16.15 $\pm$ 0.03 \\
GJ 778 & 13.431 $\pm$ 0.011 & 358.1 $\pm$ 0.3 & 2006 May & 15.34 $\pm$ 0.03 \\
GJ 778 & 12.145 $\pm$ 0.007 & 33.0 $\pm$ 0.3 & 2006 May & 16.37 $\pm$ 0.03 \\
GJ 778 & 14.466 $\pm$ 0.007 & 112.3 $\pm$ 0.3 & 2006 May & 18.18 $\pm$ 0.10 \\
GJ 778 & 16.107 $\pm$ 0.005 & 140.6 $\pm$ 0.3 & 2006 May & 17.78 $\pm$ 0.11 \\
GJ 778 & 11.575 $\pm$ 0.008 & 138.8 $\pm$ 0.3 & 2006 May & 19.48 $\pm$ 0.21 \\
GJ 778 & 10.183 $\pm$ 0.011 & 142.3 $\pm$ 0.3 & 2006 May & 17.93 $\pm$ 0.08 \\
GJ 778 & 6.633 $\pm$ 0.012 & 142.4 $\pm$ 0.3 & 2006 May & 18.42 $\pm$ 0.13 \\
GJ 778 & 10.840 $\pm$ 0.011 & 172.1 $\pm$ 0.3 & 2006 May & 16.81 $\pm$ 0.10 \\
GJ 778 & 10.369 $\pm$ 0.013 & 275.4 $\pm$ 0.3 & 2006 May & 18.36 $\pm$ 0.15 \\
GJ 778 & 10.001 $\pm$ 0.010 & 351.6 $\pm$ 0.3 & 2006 May & 17.32 $\pm$ 0.13 \\
GJ 778 & 10.691 $\pm$ 0.010 & 331.1 $\pm$ 0.3 & 2006 May & 18.97 $\pm$ 0.22 \\
GJ 778 & 11.974 $\pm$ 0.007 & 324.2 $\pm$ 0.3 & 2006 May & 19.44 $\pm$ 0.15 \\ 
\enddata
\tablecomments{All detections represent $\geq$5-sigma S/N levels.}
\tablenotetext{a}{Separation from central star.}
\tablenotetext{b}{Position angle, measured counter-clockwise from
  central star's north-south axis.}
\tablenotetext{*}{The field object described in this row is a published brown
  dwarf companion (Potter et al. 2002).  We refrained from measuring the
  magnitude of this object as blending made photometry difficult.}

\end{deluxetable}

\begin{deluxetable}{lcllll}
\tabletypesize{\scriptsize}
\tablecaption{Target Sensitivities}
\tablewidth{0pt}
\tablehead{
\colhead{}  &\colhead{} &\colhead{} &\colhead{} &\colhead{} &
\colhead{\hspace{-4.6cm}Faintest Detectable Apparent} \\
\colhead{}  &\colhead{} &\colhead{} &\colhead{} &\colhead{} &
\colhead{\hspace{-4.6cm}$K_{s}$-magnitude by Separation} \\
\colhead{Target Name} & \colhead{\hspace{0.2cm} $1.\!''$0} &
\colhead{$2.\!''$\
0} & \colhead{$3.\!''$0} & \colhead{$5.\!''$0} \\
}
\startdata
GJ 1085 & \hspace{0.2cm} 11.4 & 15.0 & 16.0 & 17.7 \\
GJ 1095 & \hspace{0.2cm} 10.1 & 13.8 & 15.4 & 17.2 \\
GJ 159 & \hspace{0.2cm} 10.7 & 13.5 & 15.1 & 16.9 \\
GJ 227 & \hspace{0.2cm} 11.3 & 14.7 & 15.7 & 17.2 \\
GJ 230 & \hspace{0.2cm} 11.9 & 15.3 & 16.6 & 17.8 \\
GJ 295 & \hspace{0.2cm} 12.2 & 15.7 & 16.6 & 17.8 \\
GJ 311 & \hspace{0.2cm} 11.7 & 15.3 & 16.3 & 17.8 \\
GJ 334.2 & \hspace{0.2cm} 11.3 & 14.7 & 16.0 & 17.4 \\
GJ 376 & \hspace{0.2cm} 11.9 & 15.3 & 16.3 & 17.8 \\
GJ 407 & \hspace{0.2cm} 11.6 & 15.0 & 16.3 & 18.0 \\
GJ 484 & \hspace{0.2cm} 11.6 & 15.0 & 16.0 & 17.5 \\
GJ 547 & \hspace{0.2cm} 11.2 & 14.7 & 15.7 & 16.3 \\
GJ 564 & \hspace{0.2cm} 12.1 & 15.3 & 16.6 & 17.8 \\
GJ 56.5 & \hspace{0.2cm} 11.6 & 15.0 & 16.0 & 17.2 \\
GJ 611 & \hspace{0.2cm} 14.0 & 17.5 & 18.2 & 18.7 \\
GJ 614 & \hspace{0.2cm} 11.9 & 15.3 & 16.6 & 17.5 \\
GJ 651 & \hspace{0.2cm} 12.5 & 15.1 & 16.1 & 17.8 \\
GJ 758 & \hspace{0.2cm} 8.0 & 11.4 & 12.7 & 14.0 \\
GJ 778 & \hspace{0.2cm} 12.5 & 15.7 & 16.9 & 17.8 \\
GJ 788 & \hspace{0.2cm} 11.9 & 15.3 & 16.3 & 17.8 \\
GJ 895.4 & \hspace{0.2cm} 11.6 & 15.7 & 16.6 & 17.8 \\
\enddata
\end{deluxetable}


\clearpage



\end{document}